\documentclass[aps,prd,reprint,amsfonts,amssymb,amsmath,showpacs,floatfix,superscriptaddress,preprintnumbers]{revtex4-2}

\usepackage{bm}
\usepackage{multirow}
\usepackage{verbatim}
\usepackage{graphicx}
\usepackage{color}

\usepackage{braket}
\usepackage{dsfont}

\usepackage{graphicx}
\usepackage{float}
\usepackage{tikz}
\usetikzlibrary{decorations.markings, positioning}

\usepackage[top=1in, bottom=1in, right=1in, left=1in]{geometry}
\usepackage{xcolor}
\usepackage{soul}
\usepackage[toc,page]{appendix}
\usepackage{subfiles}
\usepackage{listings}
\usepackage{slashed}
\usepackage[obeyspaces]{xurl}
\usepackage{hyperref}
\usepackage[most]{tcolorbox}
\usepackage{fancyhdr}
\usepackage[caption=false]{subfig}
\usepackage{booktabs}
\usepackage{amsfonts}
\usepackage{amsmath}
\usepackage{amssymb}

\definecolor{crim}{RGB}{192,64,98}
\NewDocumentCommand{\codeword}{v}{\texttt{{#1}}}
\DeclareMathAlphabet{\mathpzc}{OT1}{pzc}{m}{it}
\setulcolor{red}
\newcommand{\indspace}{\; \;}

\begin{document}
\preprint{JLAB-THY-26-4601}

\title{Accessing the Gluon Momentum Fraction of Nucleons through the Gradient Flow}
\author{Robert Edwards}
\email[e-mail: ]{edwards@jlab.org}
\affiliation{Thomas Jefferson National Accelerator Facility, Newport News, VA 23606, U.S.A.}
\author{Joe Karpie}
\email[e-mail: ]{jkarpie@jlab.org}
\affiliation{Thomas Jefferson National Accelerator Facility, Newport News, VA 23606, U.S.A.}
\author{Lorenzo Maio}
\email[e-mail: ]{Lorenzo.Maio@roma2.infn.it}
\affiliation{Dipartimento di Fisica, Universit{\`a} di Roma ``Tor Vergata" \& INFN, Sezione di Roma Tor Vergata, Via della Ricerca Scientifica 1, I-00133 Roma, Italy \\
and
Aix Marseille Univ, Universit\'e de Toulon, CNRS, CPT, Marseille, France}
\author{Christopher J.~Monahan}
\email[e-mail: ]{cjmonahan@coloradocollege.edu}
\affiliation{Department of Physics, Colorado College, Colorado Springs, CO 80903, U.S.A.}
\author{Kostas Orginos}
\email[e-mail: ]{kostas@wm.edu}
\affiliation{Department of Physics, William \& Mary, Williamsburg, VA 23187, U.S.A.}
\author{David Richards}
\email[e-mail: ]{dgr@jlab.org}
\affiliation{Thomas Jefferson National Accelerator Facility, Newport News, VA 23606, U.S.A.}
\author{Alexandru M.~Sturzu}
\email[e-mail: ]{amsturzu@wm.edu}
\affiliation{Department of Physics, William \& Mary, Williamsburg, VA 23187, U.S.A.}
\author{Savvas Zafeiropoulos}
\email[e-mail: ]{savvas.zafeiropoulos@cpt.univ-mrs.fr}
\affiliation{Aix Marseille Univ, Universit\'e de Toulon, CNRS, CPT, Marseille, France}

\collaboration{HadStruc Collaboration}
\date{\today}

\begin{abstract}
We calculate the gluon momentum fraction of the nucleon using lattice quantum chromodynamics (QCD), with a nonperturbative renormalization technique based on the gradient flow. The gluon momentum fraction is determined on a single Wilson-clover ensemble using $N_f = 2 + 1$ flavors with pion mass $358$~MeV and lattice spacing $0.094$~fm. We employ the variational method to reduce excited-state contamination and apply the distillation framework to ensure a large operator basis. To reduce systematic uncertainties, we apply Bayesian model averaging to all fit procedures. We apply matching coefficients to the flow-time dependent lattice results to recover the gluon momentum fraction in the $\overline{\mathrm{MS}}$-scheme at 2~GeV. Our final result is $\langle x \rangle_g(\mu=2\,\mathrm{GeV}) = 0.482(35)$, where we quote only statistical uncertainties. 
\end{abstract}

\maketitle

\section{Introduction}
How quarks and gluons combine to form hadrons remains a key open problem in nuclear physics. Lattice quantum chromodynamics (QCD), the numerical solution of the strong nuclear force, provides the theoretical framework that allows us to probe hadron structure from first principles and to answer questions such as: how is the momentum of a fast-moving hadron distributed amongst its constituent quarks and gluons?
\par
\begin{figure*}[t]
    \centering
    \includegraphics[width=0.9\textwidth]{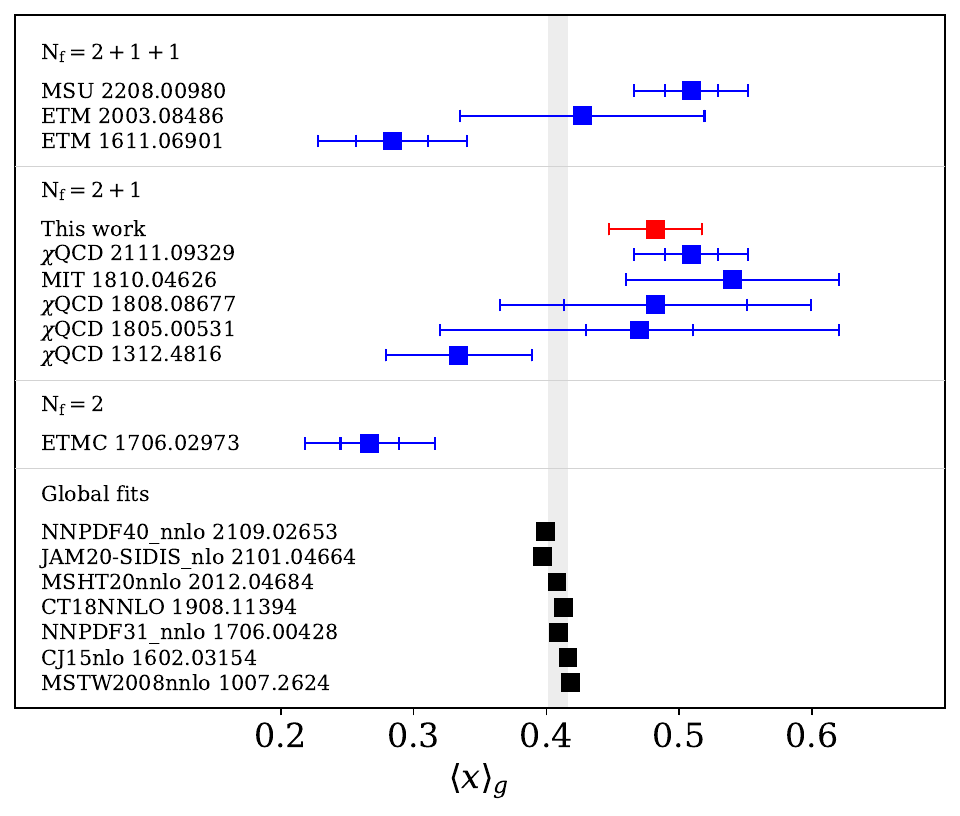} 
    \caption{Blue data points: lattice determinations of the gluon momentum fraction referenced in Table \ref{table:glu-mom-frac_latt_comparison}, including the result presented in this work in red. Black data points: determinations of the gluon momentum fraction calculated using PDF data from LHAPDF \cite{Buckley:2014ana}, and from recent JAM results \cite{Anderson:2024evk}. The grey vertical band represents the unweighted mean of these results, $\langle x\rangle_g^{\mathrm{exp.}}(\mu=2\,\mathrm{GeV}) = 0.409(7)$, to illustrate the target precision for future lattice calculations.}
    \label{fig:lattice_and_globalanalysis_glumomfrac}
\end{figure*}

 
The answer to this question is given by parton distribution functions (PDFs), which characterize the momentum carried by each of the hadron's constituents. Experimentally, PDFs are best determined by the global analysis of deep-inelastic scattering (DIS) processes \cite{Collins:1989gx}, Drell-Yan processes \cite{Collins:1982wa}, and jet production asymmetries \cite{deFlorian:2009vb}. The total momentum fraction carried by each constituent parton is captured through a weighted integral of the corresponding PDF. In general, the determinations of the momentum fractions from global analyses of experimental data are precise and generally consistent, illustrated by the data plotted as black squares in Figure \ref{fig:lattice_and_globalanalysis_glumomfrac}.
\par
From the theoretical side, the momentum fractions are typically computed using lattice QCD, the only non-perturbative, ab initio, systematically improvable approach to QCD.  Explorations of the internal structure of hadrons through lattice QCD have seen considerable progress over recent years~\cite{Monahan:2018euv, Constantinou:2020pek, Cichy:2021lih, Lin:2025hka}. Calculations focusing on the valence-quark sector are maturing and have reached sub-10\% results for PDFs in the moderate-$x$ region, although a quantitative understanding of all systematic uncertainties is incomplete~\cite{Joo:2019jct, Joo:2019bzr, Egerer:2021ymv, HadStruc:2022nay}. However, the gluon momentum fraction remains challenging to extract from lattice QCD because of the poor signal-to-noise ratio of gluon operators~\cite{Gockeler:1996zg, Gockeler:1996bm, Meyer:2007tm, Lin:2025hka}. Recent advances in algorithmic approaches and computational power have improved results~\cite{Deka:2013zha, Alexandrou:2016ekb, Alexandrou:2017oeh, Yang:2018bft, Yang:2018nqn, Alexandrou:2020sml, Wang:2021vqy, Fan:2022qve}. Recent lattice determinations for the gluon momentum fraction are summarized in Table \ref{table:glu-mom-frac_latt_comparison}.
Both lattice and global analysis results are represented in Figure \ref{fig:lattice_and_globalanalysis_glumomfrac}, which demonstrates the additional work that must be done to bring lattice calculations in line with results from phenomenology. Lattice determinations show considerable spread in central values, with significant uncertainties~\cite{Harland-Lang:2014zoa, Dulat:2015mca, Accardi:2016qay, Alekhin:2017kpj, Sato:2019yez, Hou:2019efy}. Motivated by the need for improved theoretical determinations of the gluon momentum fraction, we determine the gluon momentum fraction using several techniques to improve the signal-to-noise ratio: distillation~\cite{HadronSpectrum:2009krc}, the variational method~\cite{Edwards:2011jj}, and operator renormalization through the gradient-flow~\cite{Luscher:2009eq, Suzuki:2013gza}. 
\par 
Lattice determinations of $\langle x \rangle_g$ to date have either relied on a perturbative, one-loop renormalization of the gluon momentum fraction~\cite{Deka:2013zha, Alexandrou:2016ekb, Alexandrou:2017oeh, Alexandrou:2020sml}, or non-perturbative RI/MOM calculations of the renormalization factors~\cite{Yang:2018bft, Yang:2018nqn, Wang:2021vqy}. The use of gradient flow to renormalize calculations of PDF's and their moments have been suggested in ~\cite{Monahan:2015lha,Monahan:2016bvm,Monahan:2017hpu,Shindler:2023xpd} and performed numerically for the quark Mellin moments of the pion~\cite{Francis:2025rya, Francis:2025pgf}. In this work, we use a gauge-field operator regulated by the gradient flow. We match the flowed results to the $\overline{\mathrm{MS}}$-bar scheme via a set of matching coefficients computed in~\cite{Makino:2014taa, Harlander:2018zpi}. This approach provides a non-perturbative, gauge-invariant renormalization procedure. 
\begin{table*}
\centering
\begin{tabular}{lccccccl}\toprule
Group & ($a$[fm], $M_{\pi}$[MeV]) & $N_f$ & Fermion  & Renorm. & $\langle x \rangle_g$ \\\midrule
MSU 23 \cite{Fan:2022qve} & (0.09, 220) & 2+1+1 & Clover/HISQ & RI/MOM &    0.509(20)(23)         \\
ETMC 20 \cite{Alexandrou:2020sml}   &   (0.08, 139) & 2+1+1 &  T.M.      &     RI/MOM  &       0.427$(92)_{\mathrm{St}}$      \\
ETMC 17 \cite{ETMC2017gluonmom}   &   (0.08, 370)  & 2+1+1   &  T.M.   & NLO PT  &  0.284$(27)_{\mathrm{St}}(17)_{\mathrm{Ex}}(24)_{\mathrm{Sy}}$    \\
This work &   (0.09, 358)  & 2+1   &  Clover   &     Gradient-Flow  &  $0.482 (35)_{\mathrm{St}}$   \\
MIT 24 \cite{Hackett:2023rif}   & (0.09, 170) & 2+1 & Clover &  RI/MOM &  $0.501(27)_{\mathrm{St}}$, $0.526(31)_{\mathrm{St}}$\\
$\chi$QCD 21 \cite{Wang:2021vqy}& (0.143, 171) & 2+1 & DW  & RI/MOM &    0.509$(20)_{\mathrm{St}}(23)_{\mathrm{Sy}} $         \\
MIT 19 \cite{Shanahan:2018pib}   & (0.12, 450) & 2+1 & Clover &  RI/MOM & 0.54 $(8)_{\mathrm{St}}$\\
$\chi$QCD 18 \cite{Yang:2018bft} & (0.11, $\left\{ 135, 372 \right\}$) & 2+1 & Overlap  & RI/MOM &  0.47$(4)_{\mathrm{St}}(11)_{\mathrm{Sy}}$           \\
  & (0.14, 171) &  &  &  &             \\
 $\chi$QCD 18 \cite{Yang:2018nqn} & (0.11, 330) & 2+1 & DW &  RI/MOM      &  0.482$(69)_{\mathrm{St}}$ $(48)_{\mathrm{Sy}}$         \\
  & (0.082, 300 ) & &  &   &            \\
ETMC 17 \cite{Alexandrou:2017oeh}   &   (0.09, 130) & 2       &  T.M.      &  NLO PT   &       0.267$(22)_{\mathrm{St}}(27)_{\mathrm{Sy}}$      \\
$\chi$QCD 15 \cite{Deka:2013zha} & (0.11, $\left\{ 650, 538, 478 \right\}$) & 0 & Wilson  & NLO PT &  0.334$(55)_{\mathrm{St}}$ \\
QCDSF/UKQCD & (0.09, 665 - 1170) & 0 & Clover  & RI/MOM &  0.43$(7)_{\mathrm{St}}$$(5)_{\mathrm{Sy}}$\\
Göckeler et al.~\cite{Gockeler:1996bm} & (0.1, 585 - 993 ) & 0 & Wilson & NLO PT & 0.53(23)* \\ 
\bottomrule
\end{tabular}
\caption{Summary of lattice determinations of the gluon momentum fraction. The results reported are at a renormalization scale $\mu = 2 \;\mathrm{GeV}$ except for those with an *, which are for $\mu = 5 \; \mathrm{GeV}$. If the authors report additional details on the errors for their final results, we summarise them in the table using the subscripts ``St" for statistical, ``Ex" for experimental, and ``Sy'' for systematic.}
\label{table:glu-mom-frac_latt_comparison}
\end{table*}
\par 
The rest of this paper proceeds as follows. In Section \ref{S:GMF}, we introduce the theoretical framework for extracting gluon momentum fraction from the hadronic energy-momentum tensor. We then outline our lattice calculation in Section \ref{S:Lattice_Details}, with a more detailed description of distillation and the variational method in Section \ref{S:improvement_techniques}. We introduce the gradient flow in Section \ref{S:gf}, present results for $\langle x \rangle_g$ in Section \ref{S:results}, and summarize in Section \ref{S:conclusion}.
%
\section{Gluon Momentum Fraction} \label{S:GMF}
%
The gluon momentum fraction can be related to the $x$-weighted integral of the unpolarized gluon PDF via the first moment,
\begin{align}
    \langle x \rangle_g(\mu) = \frac{1}{2}\int^1_{-1} dx \; x\,  g(x,\mu),
\end{align} 
where $g(x,\mu)$ is the unpolarized gluon PDF at scale $\mu$. This expression is intuitively straightforward, as it corresponds to weighting the gluon PDF, a quantity related to the number of gluons at a particular momentum fraction $x$, by the momentum fraction itself. In practice, however, it is simpler to extract the gluon momentum fraction on the lattice from local operators that correspond to contributions to the energy-momentum tensor, $T^{\mu \nu}$~\cite{Collins:1981uw, ji1995qcd, Leader:2013jra}. This avoids the complications associated with extracting PDFs from lattice QCD and allows for an independent determination of the normalizing factors in the reduced gluon Ioffe-time distributions, which give access to the ratio $xg(x)/\langle x \rangle_g$~\cite{HadStruc:2021wmh}.
\par 
We begin with the decomposition of the energy-momentum tensor into the gluon and quark components as
\begin{equation}
T^{\mu \nu} = T^{\mu \nu}_g + T^{\mu \nu}_q,
\end{equation}
where
\begin{align} \label{E:SymGluonEMT}
    T^{\mu \nu}_g &= \frac{1}{g_0^2} {\rm Tr} \left\{ G^{\mu \rho} G^{\nu}_{\rho} - \frac{1}{4} g^{\mu \nu} G^{\rho \sigma} G_{\rho \sigma} \right\}, \\
    T^{\mu \nu}_q &= \overline{\psi}_f \left(\gamma^{\mu} \overleftrightarrow{D^{\nu}} + \overleftrightarrow{D^{\mu}} \gamma^{\nu} \right)\psi_f.
\end{align} 
Here the trace is over color indices and $\overset{\leftrightarrow}{D} = \overset{\rightarrow}{D} - \overset{\leftarrow}{D}$ is the symmetric covariant derivative, with $\overset{\rightarrow}{D^{\mu}} \equiv \overset{\rightarrow}{\partial^{\mu}} -igT^a A^{\mu}_a$. The gluon field strength tensor is defined as 
\begin{equation}
G^{\alpha \beta} \equiv \partial^{\alpha} A^{\beta} -\partial^{\beta} A^{\alpha} - ig \left[A^{\alpha}, A^{\beta} \right].
\end{equation} 
Note that the Lie algebra elements and their trace are implicit in the definition of the gluon operator. We use the convention $\mathrm{Tr}\left[ T^a T_b \right] = \frac{1}{2}\delta^a_{\indspace b}$.
\par 
For a nucleon state $\ket{P}$, with four-momentum $P^{\mu}$, the expectation value of the gluonic portion of the energy-momentum tensor is proportional to the gluon momentum fraction~\cite{Gockeler:1996zg, Gockeler:1996bm},
\begin{align}
    \bra{P} T^{\{\mu \nu \}}_g \ket{P} = 2 \langle x \rangle_g P^{\{  \mu} P^{ \nu \} }.
\end{align}
The curly braces denote symmetrization and trace subtraction. The nucleon states are normalized as $\braket{P|P} = 2E_n$. For the gluon field strength tensor it is useful to consider the off-diagonal components of the bilinear,  $\mathcal{O}_{1, \mu \nu} \equiv 2 G_{\mu \sigma} G_{\nu}^{\indspace \sigma} $, the trace over color degrees of freedom is now implicit. From this operator, it is useful to further define the following vector and scalar quantities,
\begin{align} \label{E:O_BOpfromEnMom}
    \mathcal{O}_{\textrm{A}i} &= \mathcal{O}_{1, i4} \nonumber\\
    \mathcal{O}_{\textrm{B}} &= \mathcal{O}_{1, 44} - \frac{1}{3}\mathcal{O}_{1, jj}.
\end{align}
These operators are proportional to the gluon momentum fraction via the relations
\begin{align} 
    \bra{P}\mathcal{O}_{\mathrm{A}i}\ket{P} &= 4iE_nP_i\langle x \rangle_g \nonumber \\
    \bra{P}\mathcal{O}_{\mathrm{B}}\ket{P} &= \left( \frac{4}{3}\boldsymbol{P}^2 + 4E_n^2 \right) \langle x \rangle_g.\label{E:O_Bop}
\end{align}
The $\mathcal{O}_B$ operator provides better signal-to-noise ratios for two reasons. First, the matrix element of the $\mathcal{O}_B$ operator can be determined in the rest frame of the nucleon, which has the largest signal-to-noise ratio. Second, the $\mathcal{O}_A$ operator is imaginary-valued, and the imaginary components of the correlation functions are typically considerably smaller in magnitude. Thus, the $\mathcal{O}_A$ operator may be more numerically unstable and noisier. We therefore follow previous lattice studies and use the $\mathcal{O}_B$ operator to extract the gluon momentum fraction from the nucleon at rest.
\section{Lattice Details} \label{S:Lattice_Details}
The gauge configurations were generated on an isotropic lattice of size $L^3 \times T = 32^3 \times 64$ with periodic boundary conditions in space using stout-link smearing in conjunction with a tree-level tadpole-improved Symanzik gauge action. The gauge configurations were updated by a rational hybrid Monte Carlo (rHMC) algorithm. A clover-improved Wilson-fermion action was used to generate $N_f =(2+1)$ dynamical flavors. Additional details of the ensemble and parameter tuning techniques can be found in Ref.~\cite{HadStruc:2021wmh} 
\par 
Three different rHMC trajectories were used to generate 1121 gauge configurations. The variance of the mean for the correlators is estimated by jackknife statistics. Within each stream, the gauge configurations were separated by 10 update steps. On each configuration, 64 temporal sources were used for the nucleon states, and 64 distillation eigenvectors were computed for each timeslice. Correlations among jackknife samples were used to estimate correlations for the fits. The field-strength tensor was implemented on the lattice using the unimproved clover operator and projected onto zero momentum. The numerical details of the ensemble are summarized in Table \ref{tab:ensemble_details}.

\begin{table}
    \centering
    \begin{tabular}{lcccccl} \toprule 
        ID & $a$ (fm) & $m_{\pi}$ (MeV) & $L^3 \times T$ & $N_{\mathrm{cfg}}$ & $N_{\mathrm{src}}$ & $N_{\mathrm{D}}$\\  
        $a094m358$   & 0.094(1)   & 358(3)   &   $32^3 \times 64$ & 1121 & 64 & 64\\  
        \bottomrule
    \end{tabular}
    \caption{Parameters of the ensemble used in this work. Here $N_{\rm cfg}$ is the number of gauge configurations used, $N_{\mathrm{src}}$ is the number of time-sources, and $N_{\rm D}$ is the rank of the distillation space. }
    \label{tab:ensemble_details}
\end{table}
\section{Improvement Techniques} \label{S:improvement_techniques}
The three-point correlation functions needed for the computation of the gluon momentum fraction have poor signal-to-noise ratio. This is primarily due to the gluonic operators. To reduce the computational cost of computing the required statistics, data improvement techniques can be used to enhance signal quality. The following section details several key techniques for improving the signal-to-noise ratio.
\subsection{Distillation} \label{SS:Distillation}
Distillation is a gauge-covariant smearing technique that, in effect, allows a factorization of correlators~\cite{HadronSpectrum:2009krc}. The pieces that factorize, called the elementals and the perambulators, correspond to hadron interpolators and propagators, respectively. These factorized pieces can be reused across different calculations. Therefore, once the initial cost of computing the elementals and perambulators is paid, subsequent calculations are significantly cheaper. This factorizable approach is particularly useful in the calculation of three-point correlation functions~\cite{Morningstar:2011ka, Lang:2014yfa, Egerer:2018xgu, Radhakrishnan:2022ubg, Batelaan:2025vhx}, even for states with large momenta~\cite{Egerer:2020hnc, Egerer:2021ymv}. Distillation clearly aids the computation of correlators, but the memory required to store perambulators is also greatly reduced relative to the full propagators. The inverse Dirac operator, $D^{-1}$, contains $(L^3 \times T \times N_c \times N_s)^2$ elements. The $N_c$ and $N_S$ terms are the number of color degrees of freedom and spinor degrees of freedom, respectively. For the ensemble used in this work, this is on the order of $10^{15}$. The propagator projected down onto distillation space, \textit{i.e.}~the perambulator, has $(N_{D} \times N_t \times N_s)^2$ elements. Here, $N_D$ is the rank of the distillation space. For the 64 distillation-eigenvectors used in this work (see Table \ref{tab:ensemble_details}), this is on the order of $10^9$. Even with point sources, this results in a significant reduction in storage requirements. 
\par 
Distillation can be particularly cost-effective when used with the variational method. Since the variational method uses a basis of operators to find the nucleon state closest to the desired (ground) state, the approach can be quite expensive due to the recalculation of propagators for every interpolator. The factorizable nature of distillation allows one to compute and reuse the same perambulators for different elementals. Therefore, the costliest aspect of the variational method can be circumvented by needing to invert the Dirac operator for only one operator in the basis of operators.
\par
To incorporate distillation, the interpolators are smeared using a Jacobi-like smearing kernel. Specifically, a low-rank approximation to the Jacobi-smearing kernel, 
\begin{align}
    J_{\sigma, n_{\sigma}} \equiv \left( 1 + \frac{\sigma \nabla^2(t)}{n_\sigma} \right)^{n_{\sigma}}\,,
    \label{E:jacobi_smearing_kernel}
\end{align}
is used to smear the quark fields. Here, $-\nabla^2(t)$ is the spatial gauge-covariant Laplacian while $\sigma$ and $n_{\sigma}$ are tunable parameters. In the limit $n_\sigma \to \infty$, the right-hand side of Eq.~\eqref{E:jacobi_smearing_kernel} approaches that of a spherically-symmetric Gaussian, and the higher energy eigenmodes of $-\nabla^2(t)$ are suppressed. The tunable parameters and the rank of the smearing kernel, $N_D$, allow one to better control this exponential suppression.  
\par
To project the Dirac matrix onto the distillation subspace, a smearing kernel constructed from the distillation subspace's basis vectors is required. The basis is constructed from the eigenvectors, $b_n$, of the gauge-covariant spatial Laplacian operator on the lattice (ordered by increasing eigenvalues). The distillation kernel is built from the outer-product of the eigenvectors,
\begin{align}
    \Box_{xy}(t) = \sum_n^{N_D} b_n(x, t) b_n^{\dagger}(y, t) \,.
\end{align}
The rank of the distillation operator $N_D$, effectively controls the width of the Gaussian smearing, which narrows as $N_D$ increases. In the limit where $\mathrm{rank}(\Box_{xy}(t)) = \mathrm{rank}(D^{-1})$, the distillation operator approaches a Dirac-delta, creating a traditional point source. In the small-rank limit, when $N_D = 1$, the distillation operator projects onto a single dimension, which is constant for the Laplacian operator, creating a traditional wall source. 
\par
\begin{figure*}
\centering
    \subfloat{%
        \includegraphics[width=0.45\textwidth]{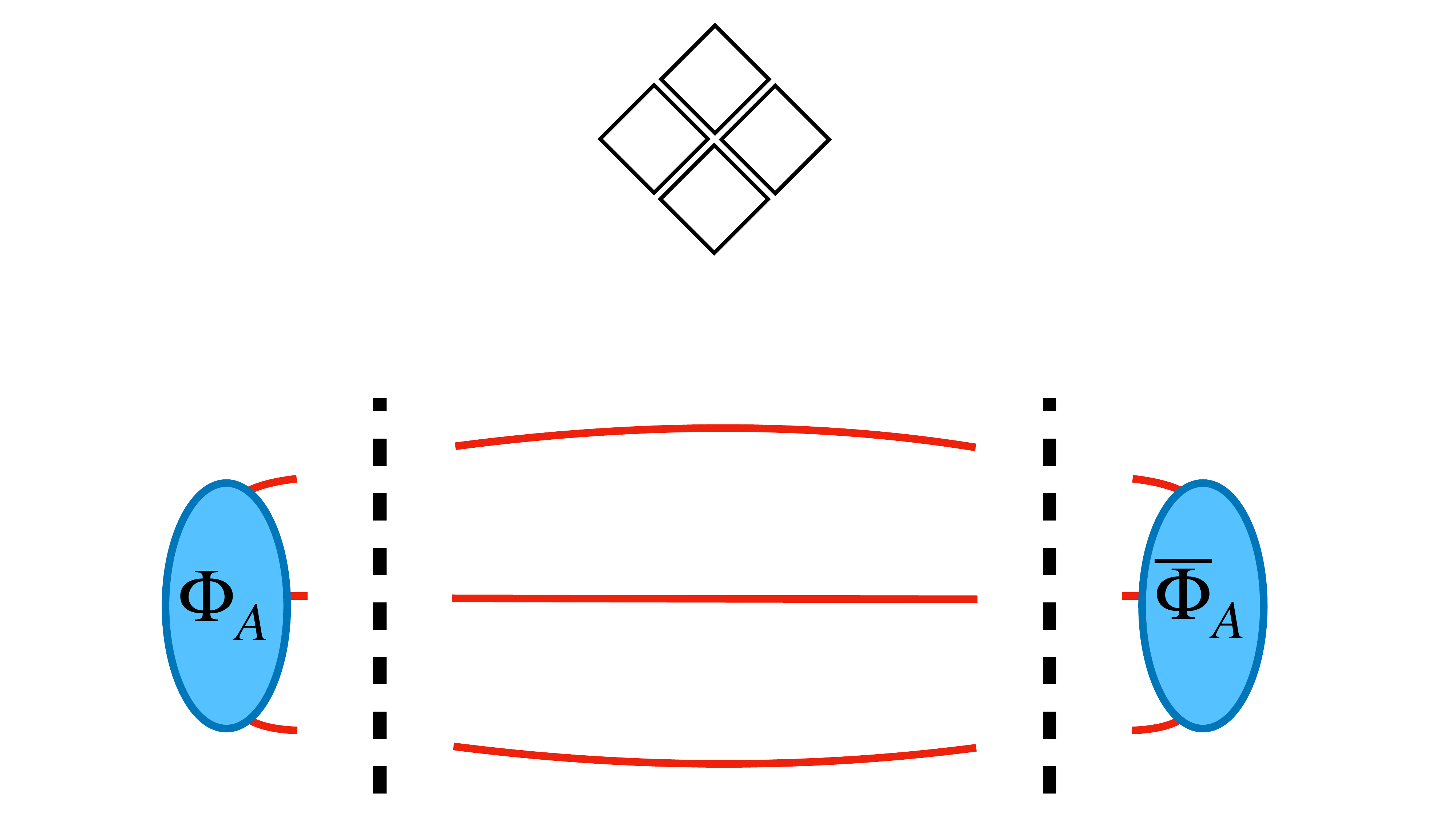}%
    }
    \hfill 
    \subfloat{%
        \includegraphics[width=0.45\textwidth]{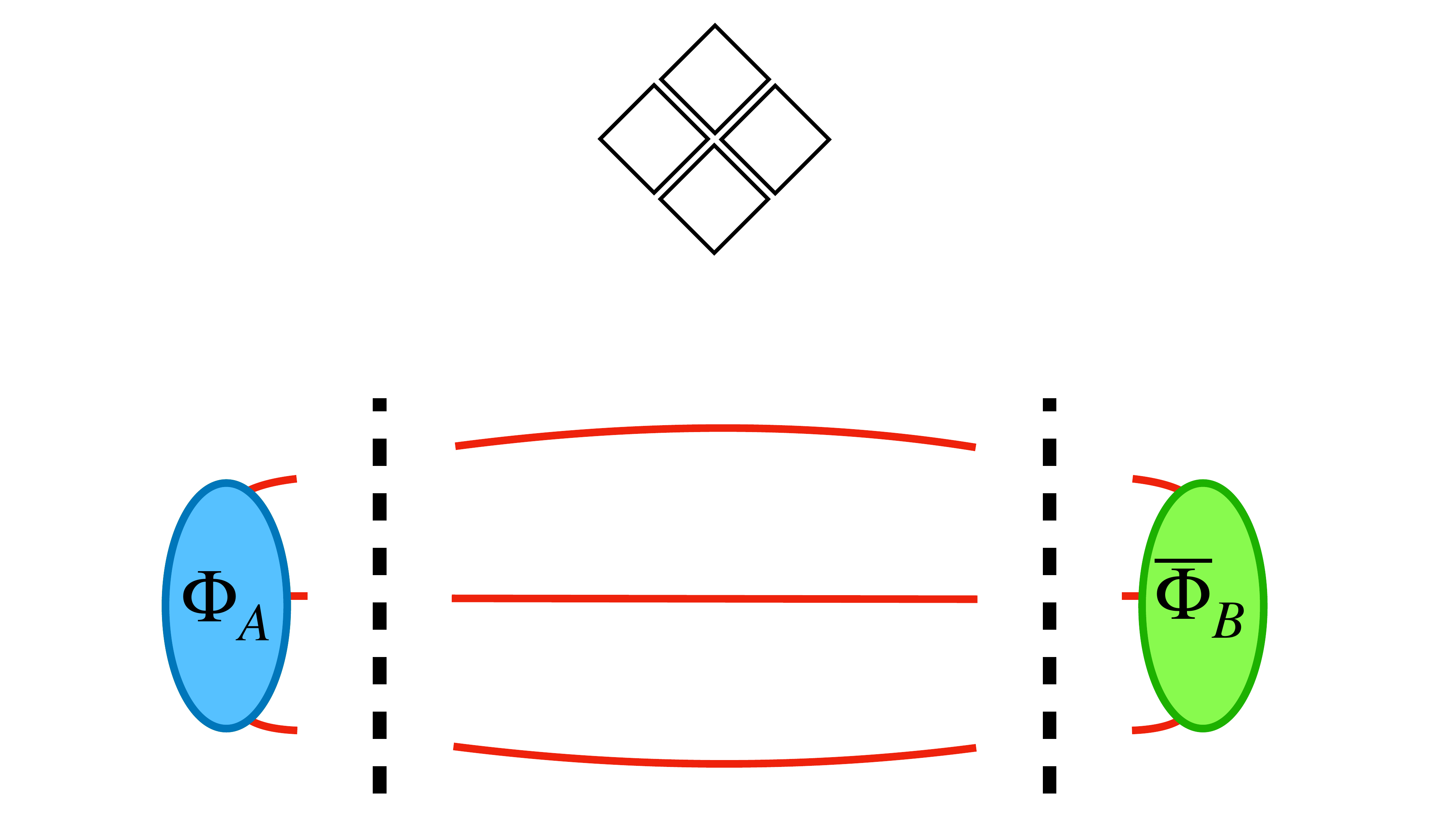}%
    }
    \caption{Two-point correlation functions under distillation. The perambulators, red lines, are factored from the elementals, the blue and green ovals. Different interpolators can be easily swapped in and out for the source/sink operators once the expensive perambulators have been computed. The gluon operator, constructed from the gauge field strength tensor, is shown as a set of four black squares.}
    \label{fig:distillation_correlation_function}
\end{figure*}

By applying the distillation operator to the quark fields, the correlation functions of hadrons can be factored into elementals and perambulators. The perambulators take the form
\begin{align}
    P_{nm}^{\alpha \beta} =  b_n^{\dagger}\left(D^{-1}\right)^{\alpha \beta}b_m.
\end{align}
The Greek indices denote spin degrees of freedom, while Latin indices are for eigenvectors. Though the perambulators account for the bulk of the computational cost, once computed, the inversions can be reused with different elements. The distilled elementals themselves contain the remaining eigenvector factors and are defined as
\begin{align}
    \Phi_{ijk}^{\alpha \beta \gamma}(t) =  \epsilon_{abc} \left( \Gamma_1 b_{i} \right)^a \left( \Gamma_2 b_{j} \right)^b
 \left( \Gamma_3 b_{k} \right)^c S^{\alpha \beta \gamma}.
\end{align}
The $\Gamma_i$ factors are combinations of gauge covariant derivatives and contain spatial and color degrees of freedom. Figure \ref{fig:distillation_correlation_function} shows the factorization of the different elementals (presented in blue and green) from the same perambulators (red), along with the disconnected gluon loops that comprise the gluon operator (black).
%
\subsection{Variational Method} \label{ss:variational_method}
\subsubsection{Two-Point Correlators}
On the lattice, the interpolators used to create desired states do not directly correspond to the expected continuum operators~\cite{Edwards:2003mv, Basak:2006ww, Dudek:2007wv, Edwards:2011jj,  Dudek:2012ag}. In general, multiple states share the quantum numbers of the operator used to construct a nucleon. These additional states include nucleon excited states and multi-particle excited states. Additionally, there is mixing between states because the $\mathrm{O}(4)$ Euclidean space-time symmetry group must be projected onto the corresponding lattice symmetry group, the rotational octahedral group $\mathrm{O}_h$. Focusing on the nucleon at rest, the spin-1/2 representations of $\mathrm{O}_h$ are typically labeled $G_{1g}$ in the literature. The subduction of the continuum symmetry representations onto the lattice symmetry representations creates a mixing between lattice operators. The operator that would typically be considered the interpolator for the nucleon in the continuum becomes contaminated with other states that possess the same quantum numbers. Regardless of the contamination source, one must prune out the undesired states. 
\par 
One way to address these contaminants is through the variational method~\cite{Creutz:1976ch, Berg:1982kp}. This approach uses a basis of interpolators, preferably those with a large overlap with the desired nucleon state, to construct a matrix of correlators, $C(t)$. This notion of overlap can be made precise within the variational method itself and is detailed below. The correlation matrix can then be diagonalized by solving a generalized eigenvalue problem (GEVP)~\cite{Blossier:2009kd},
\begin{align} \label{E:GEVP_Eqn}
    C(t)v^{\alpha}(t,t_0) = \lambda^{\alpha}(t, t_0) C(t_0)v^{\alpha}(t,t_0).
\end{align} 
The goal is to extract the generalized eigenvalues, $\lambda_i(t, t_0)$, which are referred to as principal correlators in the literature. Naturally, choosing a larger number, $N_B$, of basis operators will allow one to recover the desired state more cleanly. However, this comes at the computational cost of computing the two-point correlation functions for $N_B^2$ combinations of source-sink operators. Distillation reduces this computational cost since the elementals and perambulators are factorized. Therefore, one can swap in different interpolators, where the corresponding elementals are relatively cheap to compute, and reuse the same perambulators to avoid inverting the Dirac operator for every combination of operators.
\par 
The principal correlators are given by a tower of exponentials, $\lambda_{\alpha} = \sum_i A_{\alpha, i}e^{-E_{\alpha, i} t}$. In this work, the energies are computed by fitting the principal correlators to the constrained two-state functional form,
\begin{align} \label{E:Principle_correlator_two_state_fit}
    \lambda_{\alpha}(t) = (1 - A_{\alpha, 0})e^{-E_{\alpha}(t- t_0)} + A_{\alpha, 0}e^{-E_{\alpha}'(t - t_0)}. 
\end{align}
The ground state energy is $E_i$, while $E'$ is the energy of the first excited state. To more accurately determine the effective energy, the principal correlator should be heavily dominated by the leading exponential. Therefore, the data fit should be from time slices sufficiently separated to suppress excited-state contaminants. As is well known, however, at larger times correlators become exponentially noisier. We detail our fit procedure and data cuts for the two-point correlators in Section \ref{S:two-pt_corr}.
\par 
The set of weights needed to reconstruct the operator corresponding to a given principal correlator is given by the associated eigenvector, $v^{\alpha}(t,t_0)$. The overlap factor, $Z^\alpha_i \equiv \bra{0}\mathcal{O}_i\ket{\alpha}$ can be written in terms of the eigenvector matrix $V$ as
\begin{align}
    Z^\alpha_i = (V^{-1})^{\alpha}_i \sqrt{2m_{\alpha}}e^{m_{\alpha}t_0/2}
\end{align}
 and provides a measure of the overlap with the desired state. 
%
\subsubsection{Three-Point Correlators}
%
The three-point data is computed from the two-point correlation function and the much noisier disconnected gluon loops. The gluon operator, $\mathcal{O}_{\mathrm{B}}$, is evaluated at the insertion-time $t_g$, which can be thought of as the time-slice at which we are probing the gluonic fields of the nucleon. The three-point correlation function can be written as
\begin{align}
C_{\mathrm{3pt}}(t, t_g) &= \Big\langle \left( C_{\mathrm{2pt}}^{i}(t) - \langle C_{\mathrm{2pt}}(t) \rangle \right) \nonumber \\
&\quad \times \left( \mathcal{O}_{\mathrm{B}}^i(t_g) - \langle \mathcal{O}_{\mathrm{B}}(t_g) \rangle \right) \Big\rangle .
\label{E:Unsummed_three-point}
\end{align}
The $i$ superscript labels the configuration in the ensemble, and $\langle \dots \rangle$ is the ensemble average. The $N_{\mathrm{cfg}}$ resulting values are resampled via single-elimination jackknife. 
\par 
The variational method was used to compute the effective mass. By extending the variational method, systematic uncertainties in three-point correlation functions arising from excited-state contamination can also be reduced.  The approach is dubbed the summed-GEVP (sGEVP) method \cite{Bulava:2011yz} and is based around the summed three-point correlator, defined by summing over the insertion times of the local operator
\begin{align}
    C^{i \mathrm{, s}}_{\mathrm{3pt}}(t) = \sum^{t - 1}_{t_g = 1} C^{i}_{\mathrm{3pt}}(t, t_g).
\end{align} 
Just as for the two-point correlation functions, the tower of states created by the nucleon interpolators affects the three-point correlation functions. To isolate the terms corresponding to the ground-state nucleons, the three-point correlator on the lattice should be divided by the lattice two-point correlator. Within the variational method, this is done after rotating both the two- and three-point correlation-function matrices to the GEVP basis and selecting the desired state. The matrix element from the $i$-th configuration is then,
\begin{align}
\label{eqn:effective_mat_elem}
    \mathcal{M}^{i}&(t, t_0) \nonumber \\
    &= - \partial_t \left\{ \frac{v_n^{\dagger}\left(C^{i \mathrm{, s}}_{\mathrm{3pt}}(t) \lambda^{-1}_{i, \, 0}(t,t_0) - C^{i \mathrm{, s}}_{\mathrm{3pt}}(t_0) \right)v_n}{v_n^{\dagger}\left(C^{i}_{\mathrm{2pt}}(t_0) \right)v_n} \right\}.
\end{align}
The gluon momentum fraction on the lattice is given by the large-time limit of the ratio of three-point correlators to two-point correlators,
\begin{align} \label{E:gluon_mom_frac_final}
    \langle x \rangle_g^{\mathrm{R}} = \frac{1}{4}\frac{3}{\boldsymbol{P}^2 + 3E_0^2} \bra{P} \mathcal{O}_{\mathrm{B}}^{\mathrm{R}} \ket{P},
\end{align}
where the operator $\mathcal{O}_B^{\mathrm{R}}$ operator is renormalized. In the next section, we discuss a novel method of implementing this renormalization via the gradient flow. 

\section{Gradient Flow} \label{S:gf}
The ultraviolet divergences in the gauge fields can be removed through the gradient flow \cite{Luscher:2009eq, Luscher:2010iy, Luscher:2011bx, Luscher:2013cpa, Harlander:2025qsa}. The gradient flow embeds the fields of a Euclidean field theory with $d$ spacetime dimensions, such as lattice QCD with $d=4$, into $\mathbb{R}^d \times [0, \infty]$ space. The half-line is characterized by the flow-time parameter, $\tau$. In the continuum, the flowed gauge-fields, $B_{\mu}(\tau, x)$, are defined by a diffusion equation, with diffusion parameter  $\tau$, through
\begin{align}
    \partial_{\tau} B_{\mu} &= D_{\nu}G_{\nu \mu}, \nonumber \\
    D_{\mu} &= \partial_{\mu} + [B_{\mu}, \cdot], \\
    G_{\mu \nu} &= \partial_{\mu}B_{\nu} - \partial_{\nu} B_{\mu} + [B_{\mu},B_{\nu}]. \nonumber
\end{align}
The boundary condition for the flowed fields ensures that at vanishing flow time the physical fields are recovered, $B_{\mu}(0,x) = A_{\mu}(x)$. On a finite lattice, the existence and uniqueness of the flowed fields are guaranteed by the standard theorems of ordinary differential equations. The diffusive behavior of the above equations exponentially smooths the ultraviolet fluctuations as $\tau$ increases. Consequently, the flow-time acts as an ultraviolet regulator for composite operators \cite{Monahan:2015lha}, such as the flowed operator for the gluon momentum fraction $\widetilde{\mathcal{O}}_{i, \mu \nu}(\tau,x)$. No additional renormalization factors are required for the pure Yang-Mills energy-momentum tensor under the gradient flow. Only the spinor fields pick up a single additional multiplicative renormalization factor~\cite{Luscher:2011bx}. 
\par 
On the lattice, the flow-time equations are implemented by the equivalent condition
\begin{align}
    \partial_{\tau} B_{\mu}(\tau, x) = -g_0^2 \{ \partial_{x, \mu} \: S(B_{\mu}) \} B_{\mu}(\tau, x),
    \label{eqn:GF_on_lattice}
\end{align}
where $g_0$ is the bare coupling and $S(B_{\mu})$ is the action of the flowed fields on the lattice, which we implement with the Wilson gauge action. The derivative term is the $\mathfrak{su}(3)$-valued Lie derivative taken with respect to the link variable, $\partial_{x, \mu}f(X) \equiv T_a \partial_{x, \mu}^a f(X)$. The $T_a$ are the $\mathfrak{su}(3)$ generators and 
\begin{align}
    \partial_{x, \mu}^a f(X) = \frac{d}{d\tau} f(e^{\tau \mathfrak{g}(y , \nu)} X)\bigg|_{\tau = 0}
\end{align} 
where $\mathfrak{g}(y, \nu) = T_a$ if $(y, \nu) = (x, \mu)$ and 0 otherwise. We solve Eq.~\eqref{eqn:GF_on_lattice} on our lattices using a standard Runge-Kutta method of third order and determine the three-point correlation functions at flow times of $\tau \in \{ 0.2, 2.0\}$ in steps of 0.2 with an integration step size of 0.02.
\par 
Once an observable is computed with flowed fields, the flow-time dependence can be related to the renormalization scale $\mu$ in the $\overline{\mathrm{MS}}$-scheme using perturbation theory. This can be achieved by considering the short flow-time expansion \cite{Luscher:2011bx}, an OPE-like factorization, of the flowed operators
\begin{align}
    \widetilde{\mathcal{O}}_{i, \mu \nu}(\tau,x) = C_{ij}(\mu^2\tau)\mathcal{O}^{\mathrm{R}}_{j, \mu \nu}(\mu, x).
    \label{Eqn:short-flowtime-exp}
\end{align}
The matching coefficients, $C_{ij}(\mu^2\tau)$, have been computed to NLO in \cite{Makino:2014taa} and to NNLO in \cite{Harlander:2018zpi}.
The operators on the right-hand side are renormalized, denoted by the R superscript, $\mathcal{O}^{\mathrm{R}}_i \equiv Z_{ij}\mathcal{O}_j$ for renormalization matrix $Z_{ij}$. This linear renormalization mixes the set of twist-two operators with identical quantum numbers, including bilinear quark operators. In this work, we neglect mixing with the quark sector. Hence, we only consider mixing between the operators $\mathcal{O}_{1, \;\mu \nu}(x) \equiv G_{\mu \rho}(x)G^{ \rho}_{\; \nu}(x)$ and $\mathcal{O}_{2, \;\mu \nu}(x) \equiv \delta _{\mu \nu} G_{\rho \sigma}(x)G^{\rho \sigma}(x)$. 
In this expansion, it is implicitly assumed that the vacuum expectation values of all operators have been subtracted, 
\begin{equation}
\tilde{\mathcal{O}}_{1, \mu \nu} = G^a_{\mu \sigma} G_a^{\nu \sigma} - \langle G^a_{\mu \sigma} G_a^{\nu \sigma}  \rangle.
\end{equation}
Inverting Eq.~\eqref{Eqn:short-flowtime-exp} isolates the renormalized results
\begin{align} \label{E:Matching_equation}
    \mathcal{O}^{\mathrm{R}}_{i, \mu \nu}(\mu,x) = C^{-1}_{ij}(\mu^2\tau)\tilde{\mathcal{O}}_{j, \mu \nu}(\tau,x).
\end{align}
Therefore, by applying the inverse matching coefficients, $C_{ij}^{-1}$, to the flow-time dependent data computed on the lattice, the $\overline{\mathrm{MS}}$-bar results can be recovered. Note that the ultraviolet-finite matching coefficients $C_{ij}$ must be related to the ultraviolet-divergent matching coefficients denoted $\zeta_{ik}$ in Ref.~\cite{Harlander:2018zpi} via $C_{ij} = \zeta_{ik}Z^{-1}_{kj}$. 
%
\section{Results} \label{S:results}
%
\subsection{Two-Point Correlators} \label{S:two-pt_corr}
We implemented the variational method with an operator basis containing six interpolators. The operator basis includes both the local nucleon operator (which represents the canonical nucleon operator in the continuum and has the largest overlap with the nucleon ground state) and two hybrid interpolators built from the insertion of two covariant derivatives in an anti-symmetric combination. These hybrid operators have been shown to have large overlap with the nucleon ground state \cite{Dudek:2012ag}. The remaining operators contain different combinations of derivative and spin structures, such as the kinematically enhanced nucleon operators discussed in \cite{Zhang:2025hyo}. 
\par
\begin{figure*}
    \centering
    \includegraphics[width=0.8\textwidth]{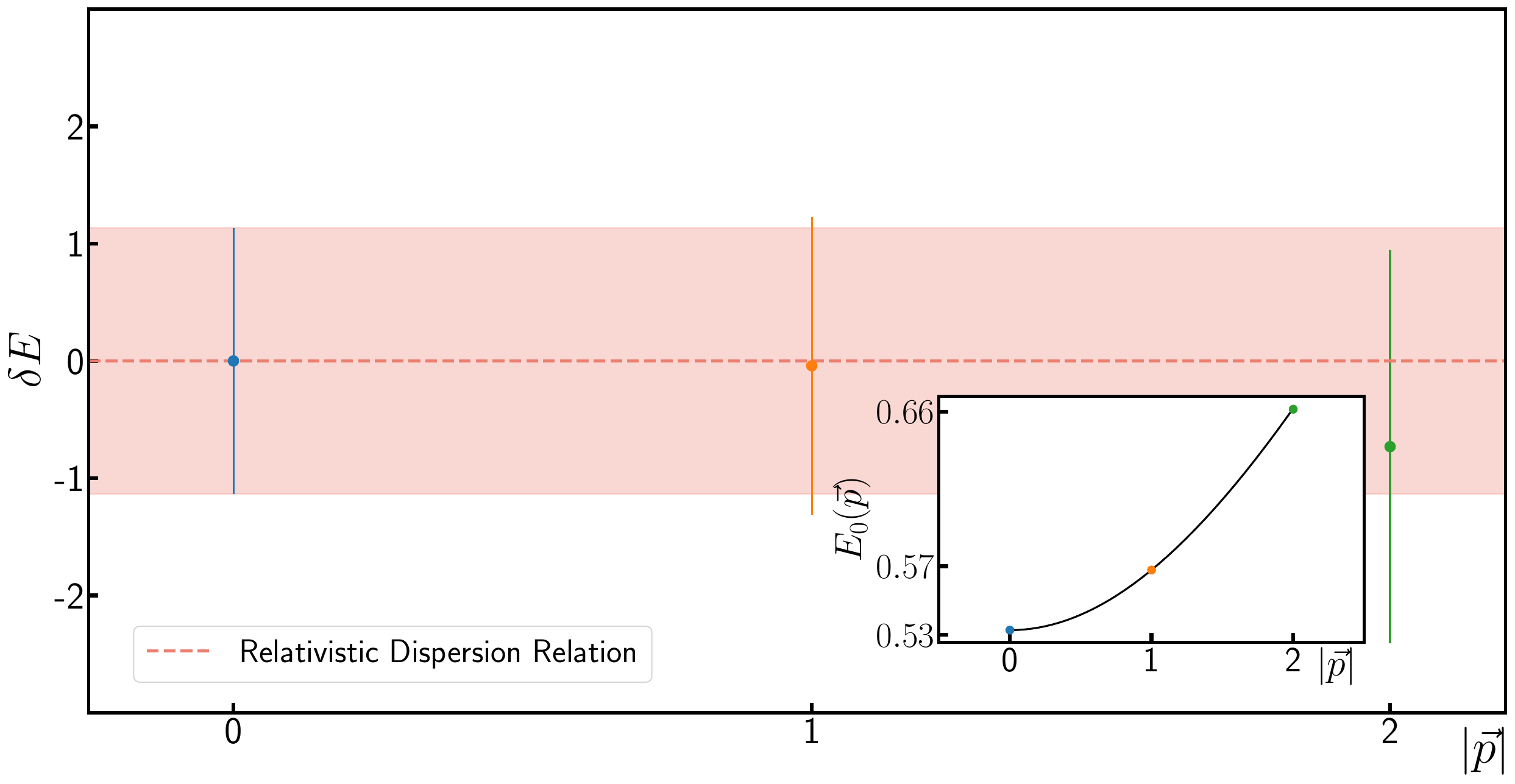}
    \caption{Analysis of the dispersion relation for the nucleon ground state. The data displayed in the main plot are defined as $\delta E \equiv \left(E_0(\vec{p}) / \sqrt{E_0(0)^2 + \vec{p}^2} - 1\right) \times 10^3$. The band around unity is set to match the uncertainty of the zero-momentum data. The inset shows the energy levels in lattice units and the analytic dispersion relation $E_0(\vec{p}) = \sqrt{E_0(0)^2 + \vec{p}^{\,2}}$, shown in black. The energies are $aE_0 \in \left[0.53281(61), 0.56781(72), 0.6614(11)\right]$.}    
    \label{fig:dispersionrelation}
\end{figure*}
\begin{figure*}
    \centering
    \includegraphics[width=0.8\textwidth]{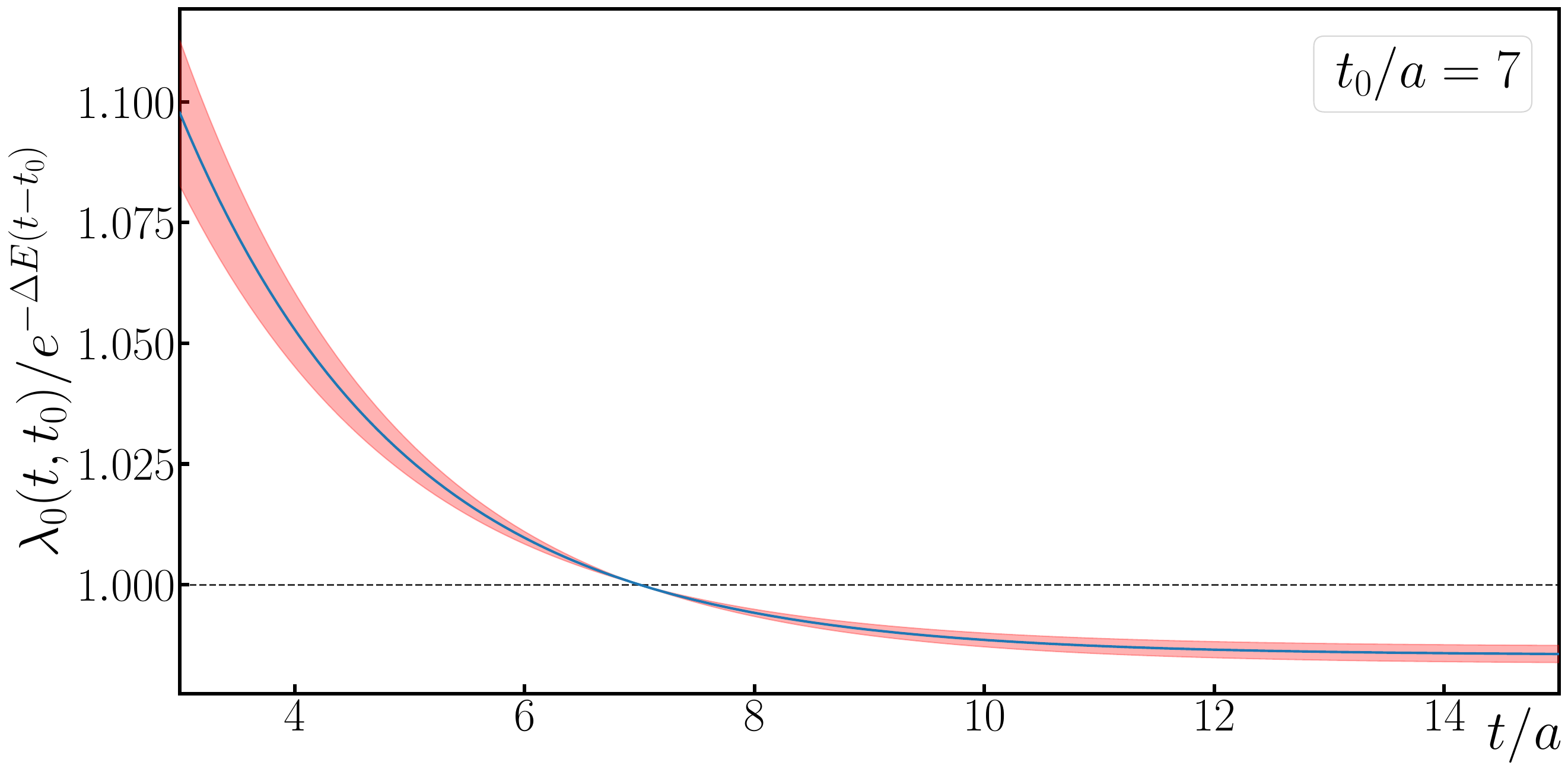}
    \caption{Principal correlators for the ground-state nucleon at rest, divided by the leading exponential. Deviations from unity characterize excited-state contaminations. The $t_0$ point is the reference time in the GEVP equation for the variational method. Therefore, it is expected that $\lambda(t_0, t_0) = 1$. The value of $t_0 = 7a$ was determined by a $\chi^2$-minimization procedure, as described further in the text.}
    \label{fig:principcorr_contaminants}
\end{figure*}

The principal correlators, $\lambda^{\alpha}(t,t_0)$, are computed from the matrix built from this basis and fit to the two-state exponential given in Eq.~\eqref{E:Principle_correlator_two_state_fit} to extract the ground state energy, $E_0$. Results for $E_0(\vec{p})$ are shown in Figure~\ref{fig:dispersionrelation}. The energies divided by the dispersion relation defined via the rest energy are shown in the main plot. The inset shows the different energies without dividing by the dispersion relation (black curve). Each single-elimination jackknife sample of the principal correlator was fitted using a Bayesian model averaging process~\cite{bayesian2020}. The covariance between data points was computed using the jackknife estimate for the ensemble. The models are defined by the number of initial data points removed from the fit. Cutting data at early times reduces the systematic uncertainties from excited-state contamination, and Bayesian model averaging reduces the model dependence of the fit results. The final result is then a weighted average, $a = \sum_M w_M a_M$, where $a_M$ are the fit results for model $M$. The weights are given by Equation 1 of Ref.~\cite{bayesian2020} and the models are defined by different functional forms or by subsets of the data. When constructing data subsets, the first three timeslices are mandatorily cut due to excited-state contamination. The stability of this choice on the final result is discussed in Section \ref{Sec:GF_matching}. The residual excited state contamination can be seen in Figure \ref{fig:principcorr_contaminants}, where the principal correlator for the nucleon is divided by the leading exponential; deviations from unity provide a measure of excited state contamination. 
\par 
As shown in Eq.~\eqref{E:GEVP_Eqn}, the principal correlators retain a dependence on the parameter $t_0$. This dependence is also evident in the tapering of the error bands around $t/a = 7$. The $t_0$ parameter should be large enough so that excited state contaminations are sufficiently suppressed so as to accurately recover the orthogonality condition $V^{\dagger} C(T_0)V = I$. However, $t_0$ cannot be too large, as noise from $C(t_0)$ will propagate into the solutions for all other time slices. To determine the optimal value of $t_0$, an augmented $\chi^2$-like quantity was proposed in \cite{Dudek:2007wv}. The distance between the original correlation matrix and the matrix reconstructed from the GEVP process, where distance is determined by the inverse of the data-correlation matrix, is used to determine the optimal value of $t_0$. Ultimately, we found this value to be $ t/a = 7$.

\subsection{Three-Point Correlators}
\begin{figure*}
    \centering
    \includegraphics[width=0.8\linewidth]{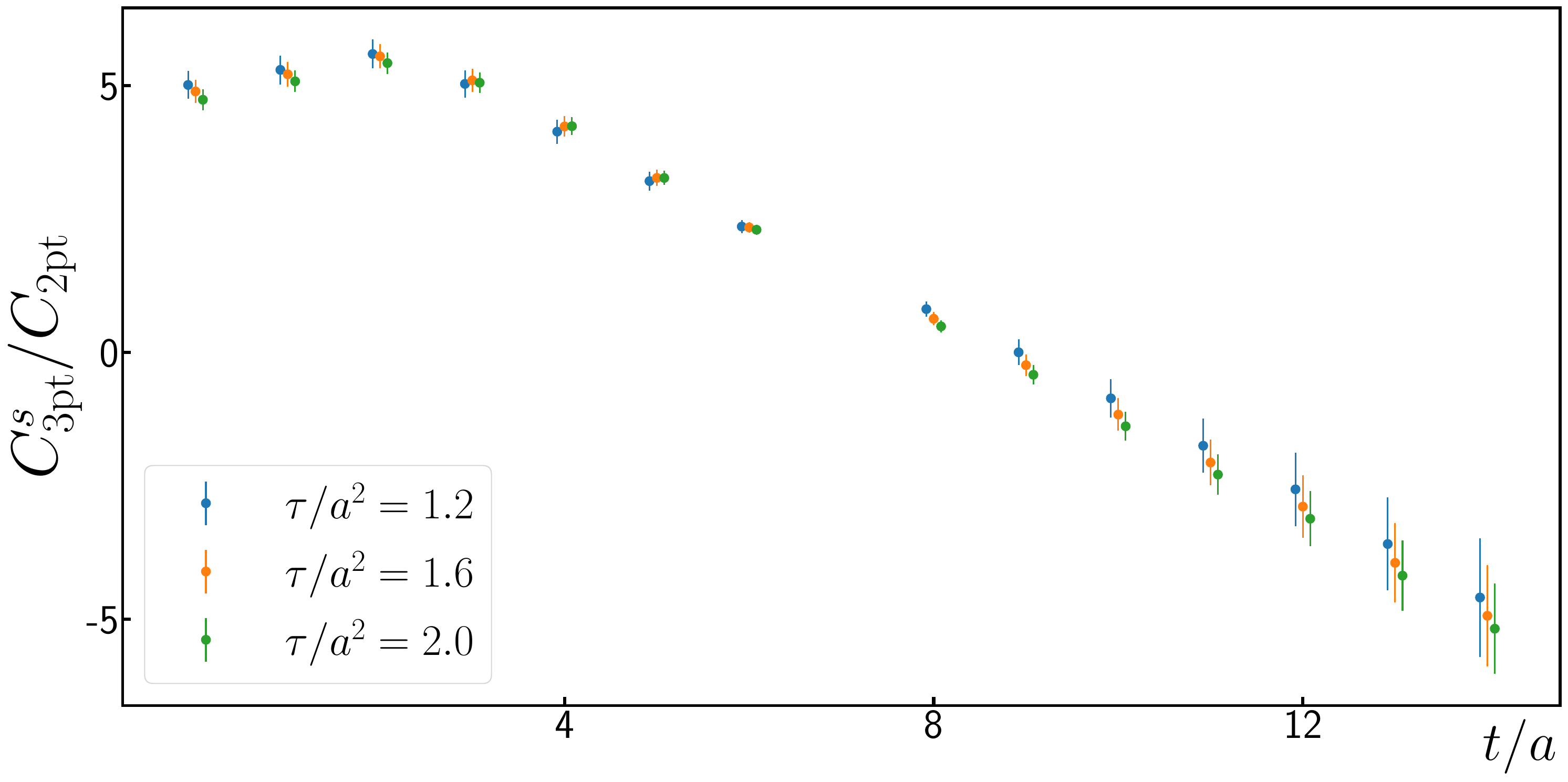}
    \caption{Ratios of the rotated and projected three- to two-point correlation matrices for three different flow-times, $\tau / a^2 \in \left\{ 1.2, 1.6, 1.8\right\}$. The data point for $t = t_0$ is made significantly noisier by the GEVP method and therefore omitted.}
    \label{fig:threept-to-twopt}
\end{figure*}

%
To numerically extract the gluon momentum function, the effective matrix element, Equation \ref{eqn:effective_mat_elem}, is fit to the functional form,
\begin{align} \label{E:matelem_funcform}
    \mathcal{M}^f(t; \tau) = A(\tau) + B(\tau) \, t \, e^{-\Delta E t}.
\end{align}
The constant term is the gluon momentum fraction, $A(\tau) = \langle x \rangle_g(\tau)$, at finite flow time $\tau$. The energy gap in the exponential is $\Delta E = E_1 - E_0$. In Figure \ref{fig:threept-to-twopt} the ratio of the summed three-point correlator to the two-point correlator is shown for two different flow times $\tau/a^2 \in \left\{1.2, 1.8 \right\}$. As expected for the disconnected diagrams computed, the statistical fluctuations are considerably more pronounced, particularly for the data at larger times, than for the two-point correlator data. 
\par
\begin{figure*}
    \centering
    \includegraphics[width=0.75\textwidth]{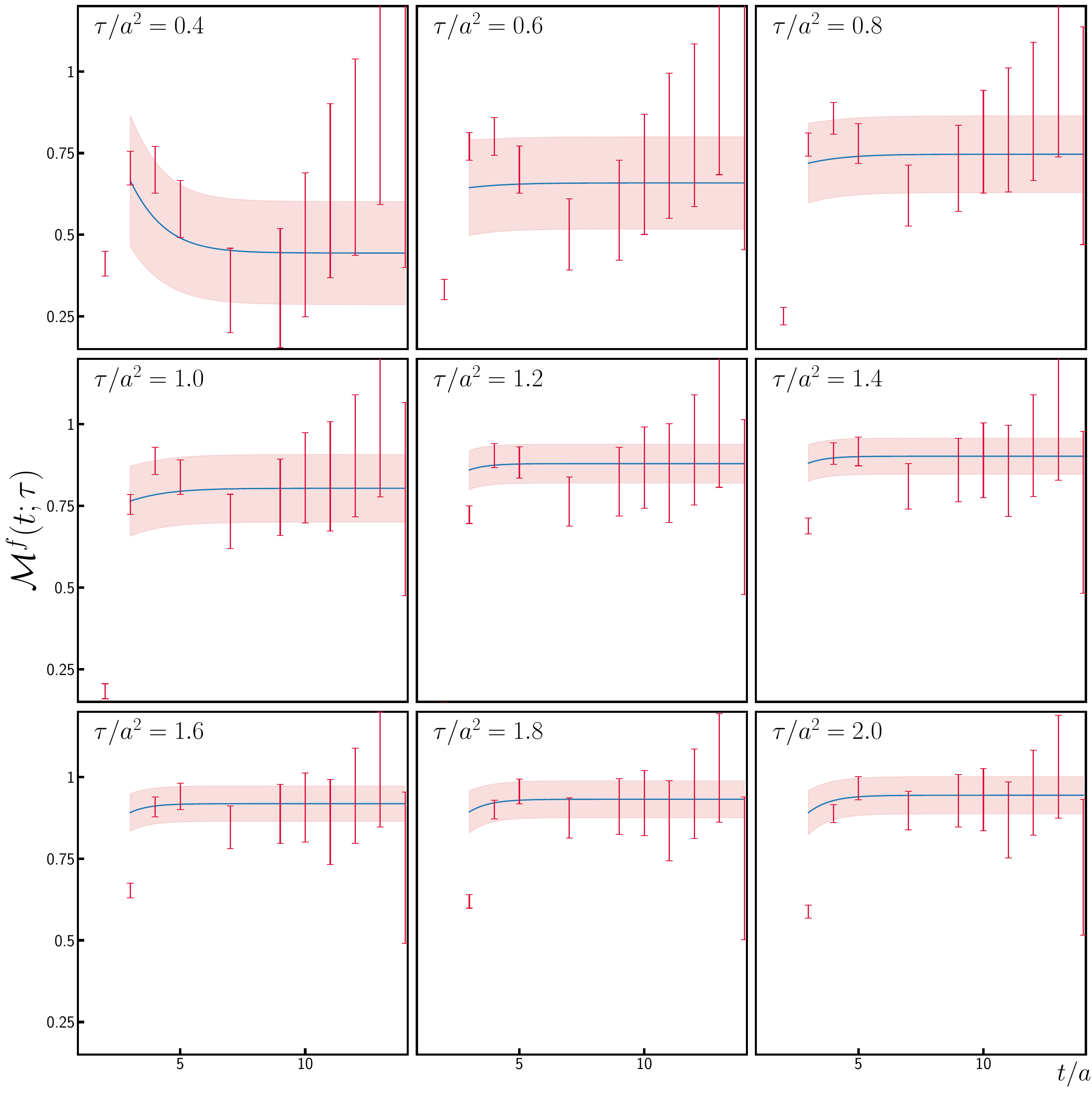}
    \caption{ Fits of the effective amplitudes for different values of $\tau$ ranging from $\tau / a^2 = 0.4$ to $\tau / a^2 = 2.0$. The momentum fraction, $\langle x \rangle_g$ is extracted from the constant (in physical time) piece of Equation \ref{E:matelem_funcform}, $A(\tau)$}
    \label{fig:eff_amp_mass_plots}
\end{figure*}

 
The final effective matrix element fit for $\tau / a^2 = 1.2$ is shown in Figure \ref{fig:eff_amp_single_plot}. The weights for the Bayesian model averaging procedure are shown below; the models are defined by the number of additional data points removed from this particular fit. Fits for a broad set of flowtimes are shown in Figure \ref{fig:eff_amp_mass_plots}. As in the two-point correlator fits, the covariance is estimated from the jackknife ensemble. Loose priors for the coefficients were used, $A = 0.75 \pm 1.0$ and $B = 1.0 \pm 2.0$, and results were found to be independent of reasonable variations in these choices of priors. The energies extracted from the two-point correlation functions was used to set a considerably tighter prior for the final parameter, $\Delta E = E_1 - E_0 \pm 2.0 * \sigma_{\Delta E}$ where $\sigma_{\Delta E}$ is the error from adding the uncorrelated uncertainties in $E_0$ and $E_1$ in quadrature. 
\begin{figure*}
    \centering
    \includegraphics[width=0.8\linewidth]{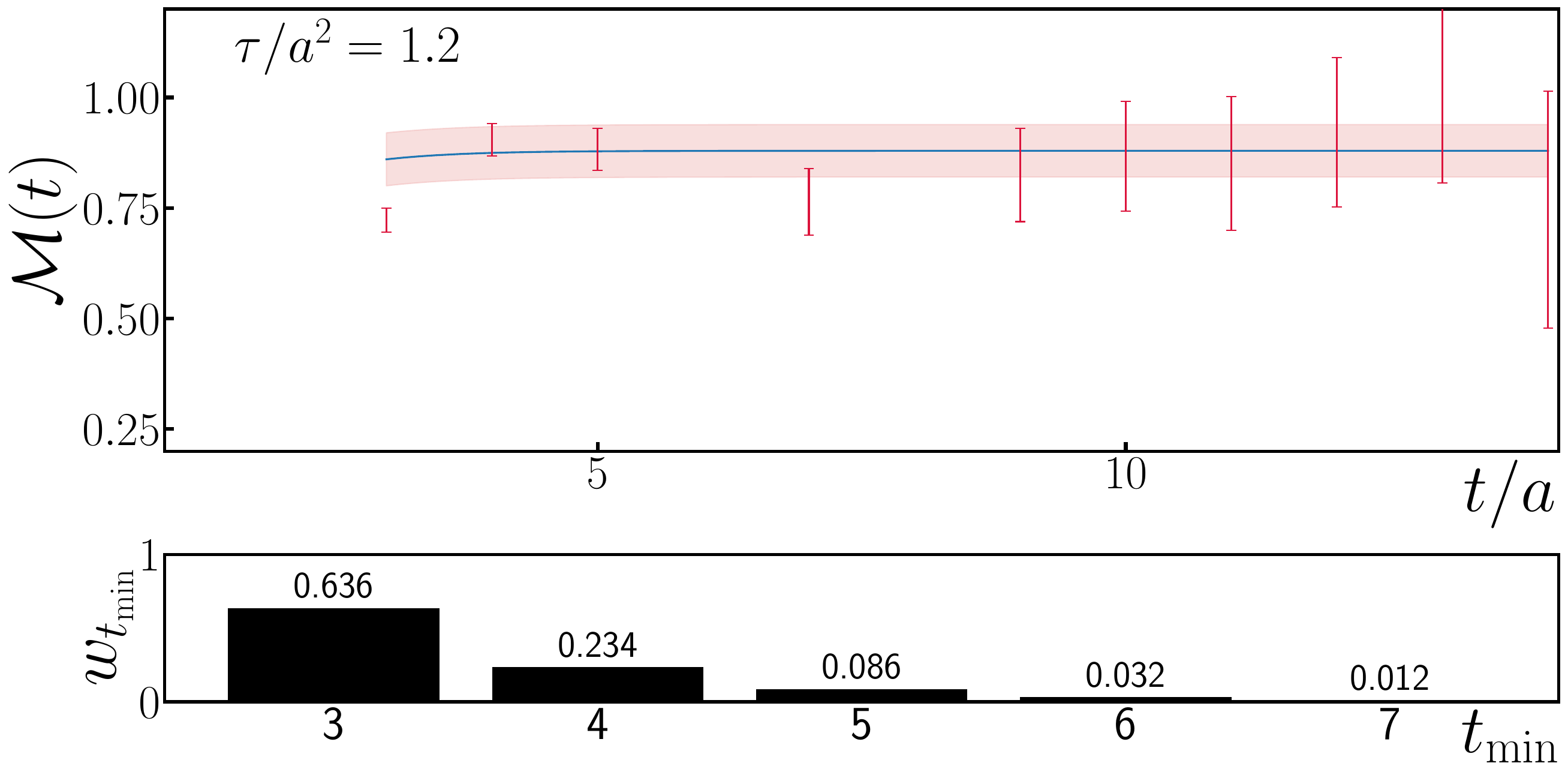}
    \caption{ Fit of the effective amplitude for the example of $\tau / a^2 = 1.2$. The lower plot shows the Bayesian model-averaging weights, with the sum of the weights equal to unity. The data corresponding to the two times flanking $t_0$, $t \in \left\{ t_0 - 1, t_0 + 1 \right\}$ are noisy, again due to the GEVP, and thus omitted. There are two points since the symmetric derivative is used. }
    \label{fig:eff_amp_single_plot}
\end{figure*}
%
\subsection{Gluon Momentum Fraction} \label{Sec:GF_matching}
\begin{figure*}
    \centering
    \includegraphics[width=0.85\textwidth]{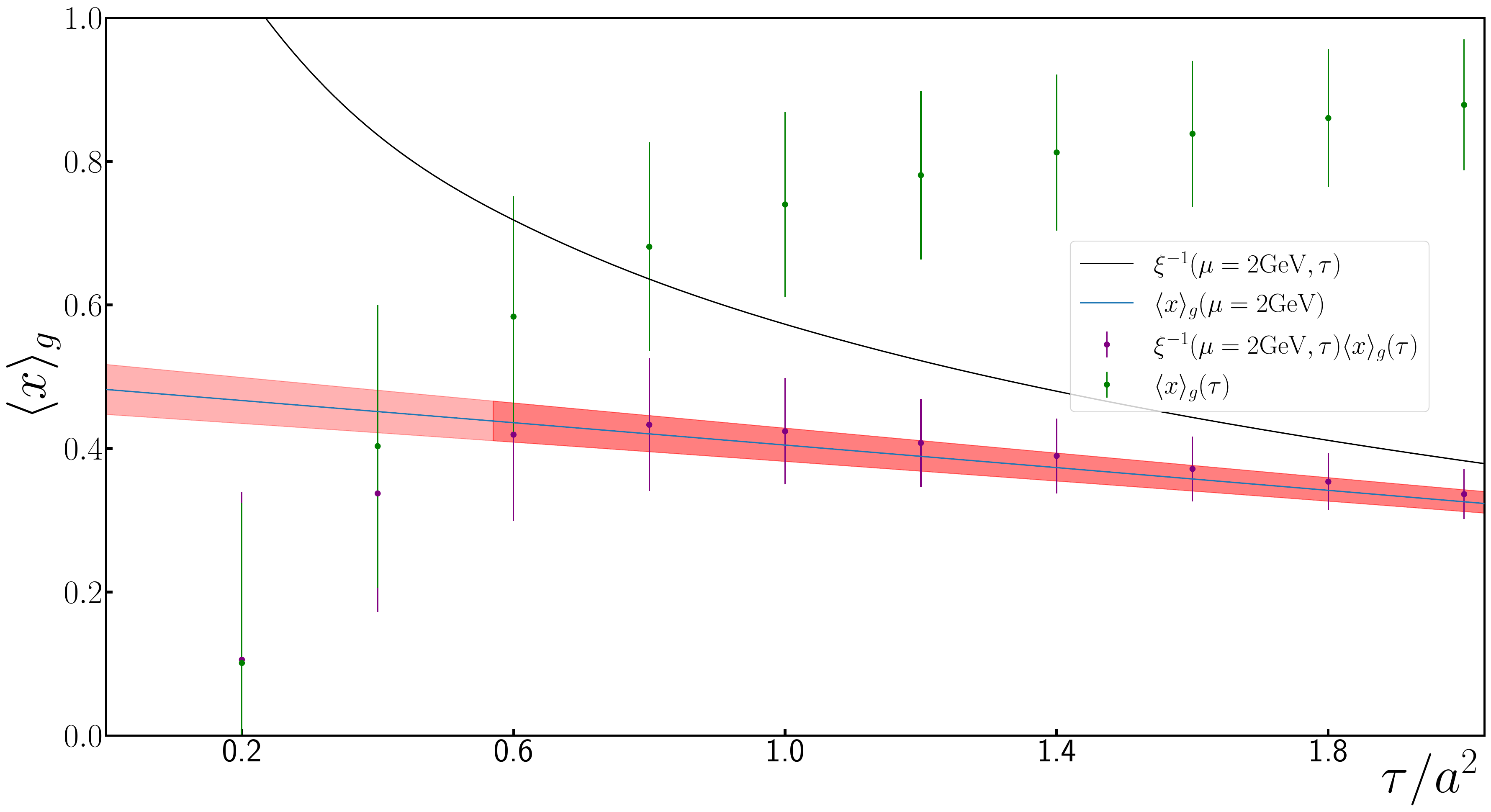}
    \caption{Results for the matching coefficients, flowed data, and matched data. The final Bayesian model-averaged fit to the flow-time dependence is also shown. The difference in shading corresponds to regions where the fit results are extrapolated beyond the smallest value for $\tau_{\mathrm{min}}/a^2 = 0.6$. Data at smaller flow times are believed to be contaminated by finite lattice-spacing effects and are therefore excluded from the fitting procedure. The final result for the gluon momentum fraction is $\langle x \rangle_g(\mu=2\,\mathrm{GeV}) = 0.482(35)$.}
    \label{fig:glumomfrac}
\end{figure*}

\begin{figure*}
    \centering
    \includegraphics[width=0.85\textwidth]{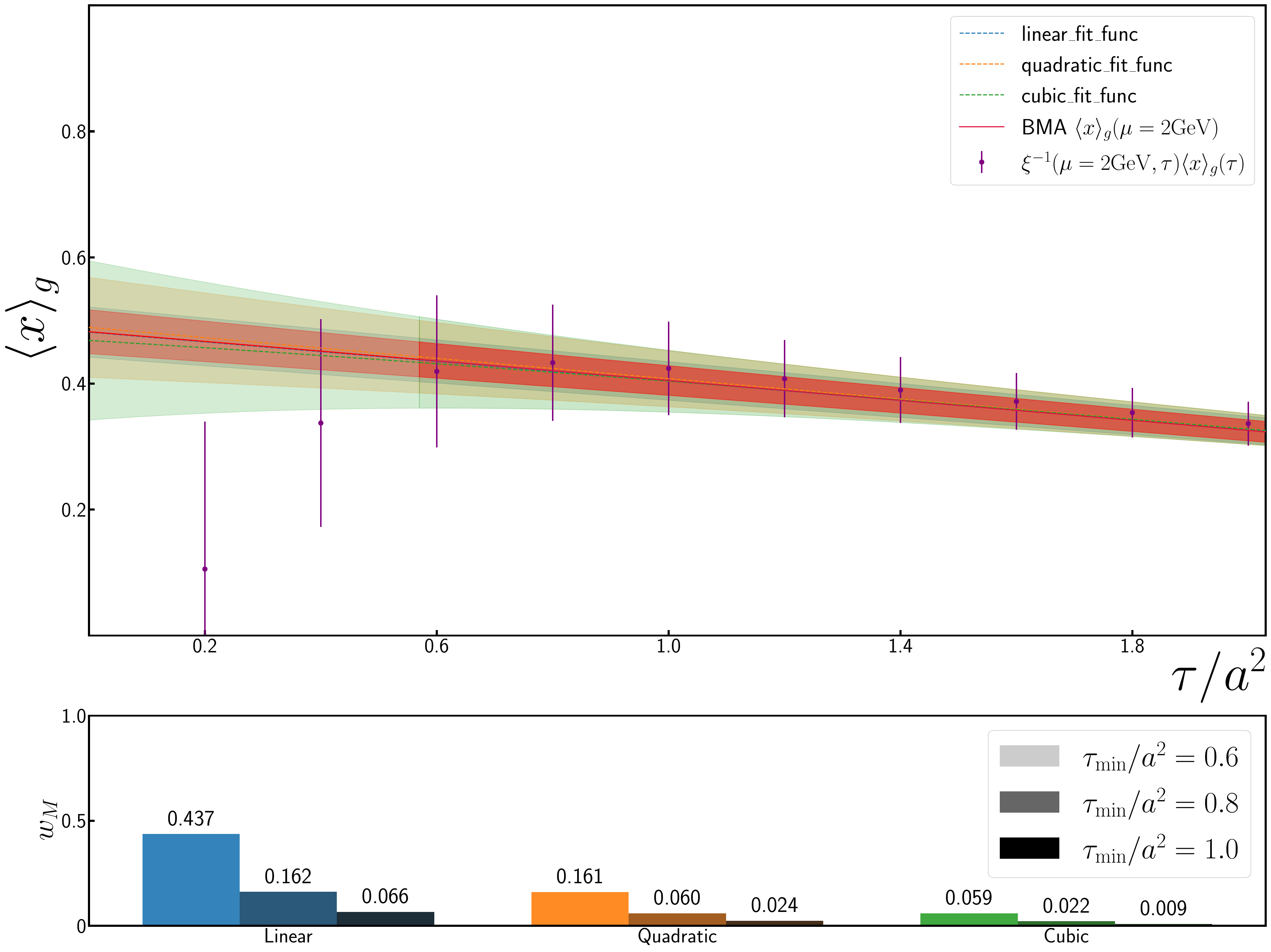}
    \caption{A focused view of the fits for the different functional forms considered in the procedure, along with the final Bayesian model-averaged result. The subplot shows a decomposition of the model weights, where darker bars correspond to larger $\tau_{\mathrm{min}}/a^2$ cuts.}
    \label{fig:glumomfrac_multifits}
\end{figure*}

%
To obtain the gluon momentum fraction in the $\overline{\mathrm{MS}}$-scheme, we apply the matching coefficients in Eqn. \ref{E:Matching_equation}. First, we fix the renormalization scale to $\mu = 2\,\mathrm{GeV}$. After applying the matching coefficients, the residual dependence of $\tau$ on the data is fit using Bayesian model averaging, with correlations across flow times included. The model averaging incorporates a set of models defined by varying the cuts at small $\tau$ and by varying the fit Ansatz to include linear, quadratic, and cubic functional forms. The application of the matching coefficients and the resulting fits are shown in Figure \ref{fig:glumomfrac}. The results for the individual models are shown in Figure \ref{fig:glumomfrac_multifits} with corresponding weights in the subplot. Our final result for the gluon momentum fraction of the nucleon, and the central result of this work, is $\langle x \rangle_g(\mu=2\,\mathrm{GeV}) = 0.482 \pm 0.035$, where we quote statistical uncertainties only.
%
\subsubsection{Systematic uncertainties}
\begin{figure}
    \centering
    \includegraphics[width=0.95\columnwidth]{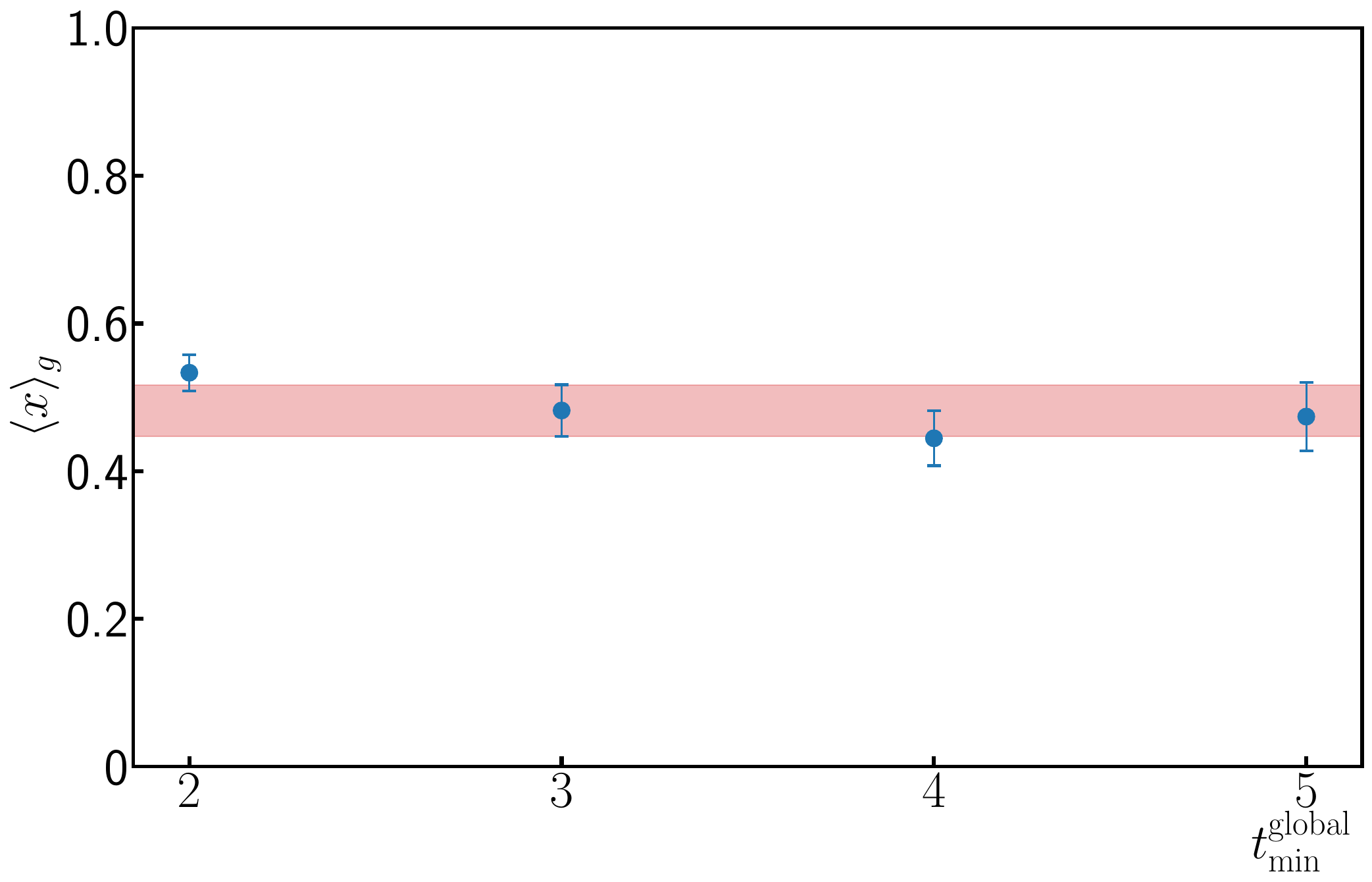}
    \caption{Stability analysis across variations in $t_{\mathrm{min}}.$ The errorband matches the result for $t_{\mathrm{min}} = 3$, our quoted final result. }
    \label{fig:stab_analysis}
\end{figure}

%
Sources of systematic uncertainties arising from excited state contamination and model dependencies have been controlled using the summed GEVP method and Bayesian model averaging. Since excited-state contamination is greatest for the early-time data, we paid special attention to determining the smallest value of $t_{\mathrm{min}}$ to use in fits involving two- and three-point correlators. The stability of the results with respect to variation in this global parameter, denoted $t_{\mathrm{min}}^{\mathrm{global}}$, is shown in Figure \ref{fig:stab_analysis}. The effects of excited-state contamination are evident by comparing the results for $t_{\mathrm{min}}^{\mathrm{global}} / a= 2$ with those for other values of $ a$. This is consistent with the observed effects of excited-state contributions in the two- and three-point functions. Therefore, we choose to use $t_{\mathrm{min}}^{\mathrm{global}} / a = 3$ for our final result. Similarly, we use a minimum value of $\tau/a^2 = 0.6$ in all fits when extrapolating the final flow-time dependence. This choice is motivated by the curvature observed at early flow times, which is attributable to lattice artifacts.
\par
To assess perturbative truncation effects, we applied two methods. First, a comparison between the one-loop (NLO), calculated in \cite{Suzuki:2013gza,Harlander:2018zpi}, and two-loop (NNLO) matching coefficients, determined in \cite{Harlander:2018zpi}, provides a broad upper bound on missing perturbative matching contributions. With NLO matching, we obtain $\langle x \rangle_g^{\mathrm{NLO}}(\mu = 2\,\mathrm{GeV}) = 0.450(32)$. The corresponding relative change is $(\langle x \rangle_g - \langle x \rangle_g^{\mathrm{NLO}} )/\langle x \rangle_g  = 0.066(95)$. This suggests that missing perturbative contributions, which will appear at NNNLO, are likely to be subdominant to our current statistical uncertainty.
\par 
As a final cross-check, we explored an alternative matching procedure, which directly relates the gluon momentum fraction at finite flow time to the $\overline{\mathrm{MS}}$ scheme at scale $\mu$. The value of the flow time is chosen such that the logarithmic terms in the matching coefficient vanish, corresponding to
$\tau = e^{-\gamma_{\mathrm{E}}}/(2\mu^2)$~\cite{Harlander:2018zpi}, where $\gamma_{\mathrm{E}}$ is the Euler-Mascheroni constant. This choice allows one to apply the matching coefficients directly to the flow-time dependent data to immediately recover $\mu$-dependent data. With this method, we obtain $\langle x \rangle_g(\mu = 2\,\mathrm{GeV}) = 0.55(6)$, which is in line with our quoted final result. However, three main difficulties arise from this approach. First, there is a strong dependence on the choice of flow time that can affect the final result. Second, this method does not take into account correlations between the $\tau$-dependent data. This means that the statistical uncertainties are not as well estimated as the main approach used above. Third, direct matching precludes estimation of polynomial contributions to the flow-time dependence, which appear in the intermediate to large flow-time regimes.
\par
We have neglected mixing with the quark sector in the application of the matching coefficients. This introduces an additional systematic effect. Numerical evidence suggests mixing with the quark sector does not significantly alter the gluon-momentum fraction \cite{Alexandrou:2016ekb, Alexandrou:2020sml}, and we defer a full analysis of quark mixing to future analyses. This analysis will enable a more comprehensive study of perturbative truncation effects. Finally, discretization effects, finite-volume effects, and unphysical pion masses also cannot be addressed with our current dataset on a single ensemble and must be studied through a full continuum, infinite-volume chiral-limit extrapolation in future work. 
\section{Conclusion \label{S:conclusion}}
In this work, the nucleon's gluon momentum fraction was computed on a single lattice ensemble using a novel renormalization approach based on gradient-flowed gauge fields. The final flow-time dependent result was matched to the $\overline{\mathrm{MS}}$-scheme via a set of perturbatively calculable coefficients. The results suggest that the gradient-flow can be an effective regulator.
\par 
The considerable statistical uncertainty arising from disconnected diagrams is controlled by using a large number of configurations, $N = 1121$. Systematic uncertainties arising from excited-state contamination or fit choices have been controlled using the summed GEVP method. Distillation enabled us to generate a large number of configurations for the six interpolators that define the basis of our GEVP method. Bayesian model averaging was used to soften the dependence on data cutoffs and functional forms.  Thus, statistical uncertainties are also well-controlled. Our final result, $\langle x \rangle_g(\mu=2\,\mathrm{GeV}) = 0.482 \pm 0.035$ is in line with current results from phenomenology, as seen in Table \ref{table:glu-mom-frac_latt_comparison}. In the absence of a complete error budget, however, this comparison must be treated carefully.
\par 
Carrying out the calculation on a single ensemble means that the final result contains finite spacing, volume, and quark-mass artifacts, which can only be addressed through future calculations on ensembles with multiple lattice spacings, volumes, and quark masses. Our result, however, suggests that gradient flow is an effective tool for determining the gluon momentum fraction in lattice QCD.
\section{Acknowledgments}
We would like to thank the members of the HadStruc Collaboration.
This material is based upon work supported by the U.S. Department of Energy, Office of Science, Office of Workforce Development for Teachers and Scientists, Office of Science Graduate Student Research (SCGSR) program. The SCGSR program is administered by the Oak Ridge Institute for Science and Education for the DOE under contract number DE-SC0014664. L. M. was supported by an Emergence grant of CNRS Physics. 
A.M.S.~and C.J.M.~are supported in part by U.S.~DOE Grant \mbox{\#DE-SC0025908} and in part by U.S.~DOE ECA \mbox{\#DE-SC0023047}. A.M.S.~is also supported in part by the Quark-Gluon Tomography (QGT) Topical Collaboration, U.S.~DOE Award \mbox{\#DE-SC0023646}. The work of K.O. was supported in part by U.S. DOE Grant No. DE-FG02-04ER41302. The work of D.R.\ was conducted in
part under the Laboratory Directed Research and Development Program at
Thomas Jefferson National Accelerator Facility for the U.S. Department of Energy. 
SZ acknowledges support in part from the Agence Nationale de la Recherche (ANR) under Project No. ANR-23-CE31-0019. R.E., J.K., and D.R. acknowledges support from the U.S. Department of Energy (DOE), Office of Science, under Contract No. DE-AC05-06OR23177, under which Jefferson Science Associates, LLC operates Jefferson Lab. This paper is based upon work supported by the U.S. Department of Energy, Office of Science, Office of Advanced Scientific Computing Research project American Science Cloud (AmSC) -- part of the Genesis Mission.

The software packages \texttt{lsqfit} ~\cite{peter_lepage_2025_15874755}, \texttt{gvar} \cite{peter_lepage_2025_15874477},
{\tt Chroma}~\cite{Edwards:2004sx}, {\tt QUDA}~\cite{Clark:2009wm,Babich:2010mu}, {\tt QUDA-MG}~\cite{Clark:2016rdz}, {\tt QPhiX}~\cite{Joo:2013lwm}, {\tt MG\_PROTO}~\cite{MGProtoDownload}, 
and {\tt Redstar}~\cite{Chen:2023zyy} were used. 
Some software codes used in this project were developed with support from the U.S.\ Department of Energy, Office of Science, Office of Advanced Scientific Computing Research and Office of Nuclear Physics, Scientific Discovery through Advanced Computing (SciDAC) program; also acknowledged is support from the Exascale Computing Project (17-SC-20-SC), a collaborative effort of the U.S.\ Department of Energy Office of Science and the National Nuclear Security Administration.

This work used clusters at Jefferson Laboratory under the USQCD Initiative and the LQCD ARRA project.
Also used was an award of computer time provided by the U.S. Department of Energy INCITE program and supported in part under an ALCC award, and resources at: the Oak
Ridge Leadership Computing Facility (OLCF), which is a DOE
Office of Science User Facility supported under Contract
DE-AC05-00OR22725; the National Energy Research Scientific Computing Center (NERSC), a U.S. Department of Energy Office of Science User Facility located at Lawrence Berkeley National Laboratory, operated under Contract No. DE-AC02-05CH11231; the Texas Advanced Computing Center (TACC) at The University of Texas at Austin; the Extreme Science and Engineering Discovery Environment (XSEDE), which is supported by National Science Foundation Grant No. ACI-154856; and Argonne Leadership Computing Facility (ALCF), a U.S. DOE Office of Science user facility at Argonne National Laboratory, which is supported by the Office of Science of the U.S. DOE under Contract No. DE-AC02-06CH11357. This work also benefited from access to the Jean Zay supercomputer at the Institute for Development and Resources in Intensive Scientific Computing (IDRIS) in Orsay, France, under project 2025-A0180516207.

\section*{Data availability}
The data that support the findings of this study are openly available in Zenodo under the Ref. \cite{Edwards2026GluonDataset}.

\bibliography{References}

\appendix

\end{document}